\documentclass[12pt,twocolumn]{iopart}
\usepackage{times,graphics,graphicx,bm}
\usepackage{amsthm}
\usepackage{amssymb}
\usepackage{iopams}
\begin{document}


\title[Procedure for measuring the $g^{(2)}(0)$ of a CW telecom heralded single-photon source]{Towards a standard procedure for the measurement of the multi-photon component in a CW telecom heralded single-photon source}


\author{E. Rebufello$^{1,2}$, F. Piacentini$^1$, M. L\'{o}pez$^3$, R. A. Kirkwood$^4$, I. Ruo Berchera$^1$, M. Gramegna$^1$, G. Brida$^1$, S. K\"{u}ck$^3$, C. J. Chunnilall$^4$,  M. Genovese$^1$, I. P. Degiovanni$^1$}
\address{$^1$ Istituto Nazionale di Ricerca Metrologica (INRiM), Torino, Italy}
\address{$^2$ Politecnico di Torino, Torino, Italy}
\address{$^3$ Physikalisch-Technische Bundesanstalt (PTB), Braunschweig and Berlin, Germany}
\address{$^4$ National Physical Laboratory (NPL), Teddington, UK}
\ead{f.piacentini@inrim.it}
\vspace{10pt}

\begin{abstract}
Single-photon sources are set to be a fundamental tool for metrological applications as well as for quantum information related technologies.
Because of their upcoming widespread dissemination, the need for their characterization and standardization is becoming of the utmost relevance.
Here, we illustrate a strategy to provide a quantitative estimate of the multi-photon component of a single-photon source, showing the results achieved in a pilot study for the measurement of the second order autocorrelation function $g^{(2)}$ of a low-noise CW heralded single photon source prototype (operating at telecom wavelength $\lambda=1550$ nm) realized in INRiM.
The results of this pilot study, involving PTB, NPL and INRiM, will help to build up a robust and unambiguous procedure for the characterization of the emission of a single-photon source.

\end{abstract}

%
%
%
%
%

\section{Introduction}

Single-photon sources (SPSs) are crucial for metrological \cite{metrology_3,metrology_1,metrology_2,tarates,metrology_4}, fundamental \cite{brunner,alicki_2,alicki_3,alicki_1,deraedt1,pusey,sequential,leggett-garg,protective,scirep} and applied research \cite{quantumcomm,quantumtech,bana,eliz}.\\
An ideal single-photon source should exhibit the features of deterministic behavior (single photons can be emitted arbitrarily by the user with $100\%$ probability and with the highest possible repetition rate), indistinguishability among the photons emitted at different times, and zero multi-photon component \cite{shell_2,alan,zbin}.
Although such a device is still far from being realized, much effort and many different ideas have been invested in these sources and their improvement \cite{shell_2,alan,zbin,dema,ursin,silber,villo,marq}.\\
After an initial attempt aiming primarily to enhance the rate and efficiency of the single-photon emission, nowadays many teams worldwide prefer to focus on designing specific on-purpose SPS prototypes, improving the feature(s) needed for the related application(s).
For example, the deterministic photon emission could be in principle achieved with single-photon emitters like color centers in diamond \cite{diam1,diam2,pino1,pino2}, quantum dots \cite{dot1,dot2,dot3}, atomic ensembles \cite{atom}, single ions \cite{ion1,ion2} and molecules \cite{mol1,mol2}.
Nevertheless, such systems are always affected by a non-unit photon extraction efficiency, a drawback that practically blurs their deterministic behavior with respect to intrinsically probabilistic heralded single-photon sources (HSPSs), such as those based on four-wave mixing \cite{fwm1,fwm2} and parametric down-conversion (PDC) \cite{alan,ursin,silber,marq,HSPS1,HSPS2,HSPS3}.\\
Single-photon technologies are nowadays of the utmost relevance in several research branches, and even for upcoming commercial implementations (e.g. secure communication); this makes their characterization a crucial metrological task.
The multi-photon component in SPSs emission is one of the main issues of such devices, and so a standard methodology to properly quantify such component would be of major interest for a large scientific community.\\
In a joint effort to provide a robust procedure suited for this task, three metrological institutes (INRiM, NPL and PTB) have participated in a measurement campaign addressed to the measurement of the multi-photon emission of a particular prototype of PDC-based CW HSPS \cite{NHSPS,NHSPS2}, characterized by a very low number of residual non-heralded ``noise'' photons granting a small multi-photon output.
This research aims to prepare a robust procedure for an international metrological comparison on the $g^{(2)}(0)$ measurements of extremely faint light CW sources (i.e. at the single photon level) in the telecom bandwith.\\
An analogous effort has been addressed by the same team to establish a proper procedure for the measurement of the $g^{(2)}$ function of a single-emitter pulsed SPS (based on nitrogen vacancies in diamonds emitting in the visible range); the related results can be found in \cite{pino_sps}.\\
These joint efforts will be crucial for the development of an international metrological infrastructure for the characterization of both CW and pulsed SPSs, paving the way to the commercial success of the forthcoming quantum photonics related technologies.\\
In the following, the results achieved for the CW telecom HSPS will be presented.\\


\section{The single-photon source}

\begin{figure}[t]
\begin{center}
\includegraphics[width=\textwidth]{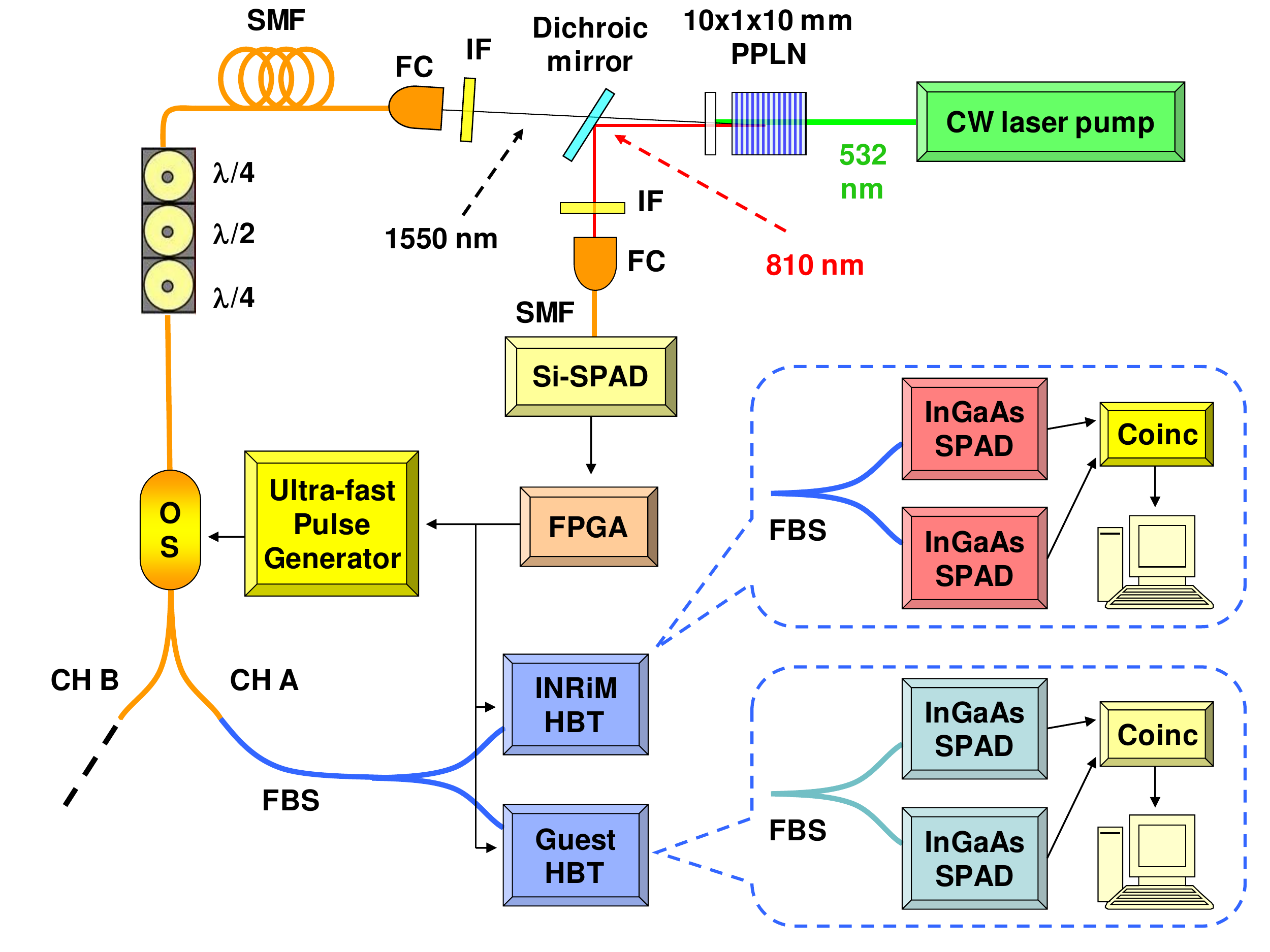}
\caption{Experimental setup. PPLN: periodically-poled lithium niobate. FC: fiber coupler. SMF: single-mode fiber. $\lambda/4$: quarter-wave fiber paddle. $\lambda/2$: half-wave fiber paddle. OS: optical shutter. FPGA: field programmable gate array. FBS: fiber beam splitter. HBT: Hanbury-Brown \& Twiss interferometer. Coinc: time-tagging coincidence electronics.
}
\label{setup}
\end{center}
\end{figure}

In our experimental setup (Fig.\ref{setup}) a continuous wave (CW) laser ($\lambda=532$ nm) pumps a $10\times1\times10$ mm periodically-poled Lithium Niobate (PPLN) crystal, producing non-degenerate PDC.
We choose signal and idler photons with wavelengths of $\lambda_{s}=1550$ nm and $\lambda_{i}=810$ nm, respectively.
The idler photon is sent to an interference filter (IF) with a full width at half maximum (FWHM) of 10 nm, then fiber-coupled and sent to a Silicon single-photon avalanche detector (Si-SPAD), heralding the arrival of a 1550 nm signal photon.
The signal photon is addressed to a 30 nm FWHM IF and coupled to a 20 m long single-mode optical fiber connected to an electro-optical shutter (OS) operated by a fast pulse generator controlled by a field programmable gate array (FPGA).
As an optical shutter we use an EO-space Ultra-High-Speed $2\times2$, polarization-dependent optical switch, 
whose technology is based on a LiNbO$_3$ Mach-Zehnder interferometer.\\
On receipt of a heralding signal, the FPGA triggers a pulse generator that opens our HSPS output channel, i.e. OS channel A, for a time interval $\Delta t_{\textrm{switch}}=7$ ns which corresponds to the passage of a 1550 nm photon, and then switches to channel B for a chosen minimum ``sleep'' time $t_{\textrm{dead}}\simeq11$ $\mu$s before accepting a new heralding.
This way we can adjust the rate at which single photons are emitted from our device, granting a minimum time between subsequent photons, thus avoiding dead time issues with many detectors.\\
For the purpose of this joint measurement, the HSPS output is connected to a 50:50 fiber beam splitter (FBS) whose outputs are sent to two Hanbury Brown \& Twiss (HBT) interferometers, one belonging to INRiM and the other to the guest NMI (PTB or NPL).
This configuration allows the performance of simultaneous data collection between INRiM and the guest NMI, avoiding any mismatch between measurements because of some drift in the HSPS output over time.
This way, we have two joint measurement sessions, one involving INRiM and NPL (session INRiM-NPL) and one with INRiM and PTB (session INRiM-PTB).\\
Every HBT involved is composed of two infrared InGaAs-InP SPADs, be they free-running or gated. In gated mode, the SPADs are triggered by the same FPGA signal that goes to the OS.
In each HBT, the outputs of the two InGaAs SPADs are sent to time-tagging coincidence electronics.\\

\section{The measurement}

\begin{figure}
\begin{center}
\includegraphics[width=\columnwidth]{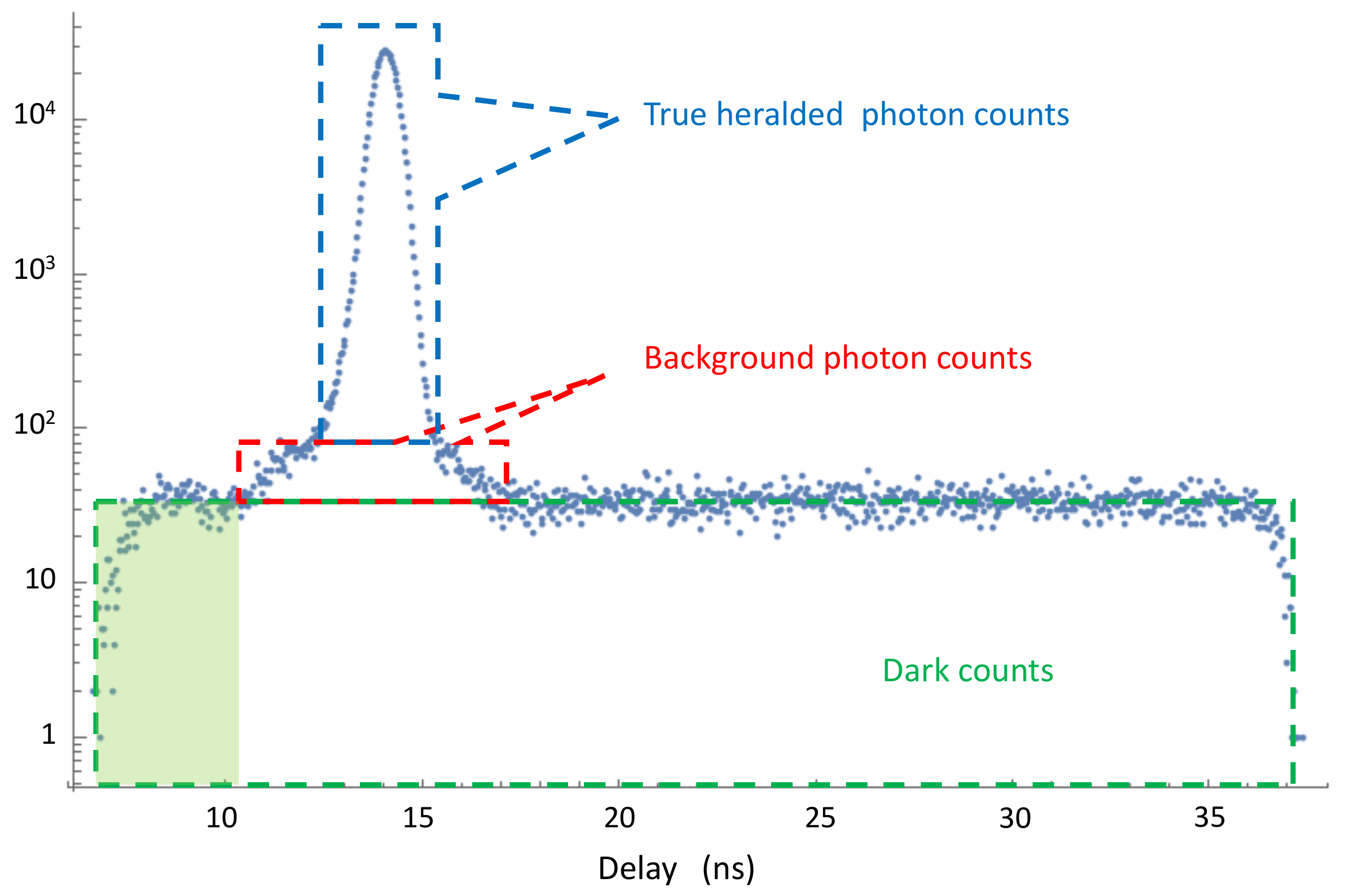}
\caption{Histogram of the detections acquired in one run by one of the INRiM HBT gated SPADs, with a 30 ns detection window. The true heralded photons (Her), background photons (Bkg) and dark count (Dark) contributions can be clearly seen. The green shaded area indicates the dark counts forbidding the SPAD to detect a photon from the HSPS.} \label{peak-in}
\end{center}
\end{figure}
Looking at the temporal histogram of one of INRiM HBT gated SPADs (Fig.\ref{peak-in}), we can distinguish three different ``regions'' corresponding to:
\begin{itemize}
    \item $N^\mathrm{(Her)}$ = true heralded photon counts;
    \item $N^\mathrm{(Bkg)}$ = counts due to unheralded background photons passing
    through the optical switch;
    \item $N^\mathrm{(Dark)}$ = dark counts of the InGaAs SPAD.
\end{itemize}
We define the total photon detection probability (as seen by the HBTs SPADs) for each trigger count as:
\begin{equation}\label{P-T}
P_{i}^{\mathrm{(Ph)}} \equiv  \frac{N_{i}^{\mathrm{(Ph)}}}{N_{i}^{\mathrm{(Trig)}}} = \frac{N_{i}^{\mathrm{(Her)}}+N_{i}^{\mathrm{(Bkg)}}}{N_{i}^{\mathrm{(Trig)}}} = \frac{N_{i}^{\mathrm{(Tot)}}-N_{i}^{\mathrm{(Dark)}}}{N_{i}^{\mathrm{(Trig)}}}~~~~~~~~~~~~~~i=1,2
\end{equation}
where $N_{i}^{\mathrm{(Trig)}}$ is the total number of trigger counts accepted by the $i$-th detector and $P_{i}^{\mathrm{(Tot)}}=P_{i}^{\mathrm{(Her)}}+P_{i}^{\mathrm{(Bkg)}}+P_{i}^{\mathrm{(Dark)}}$ is the overall detection probability of that SPAD.
By carefully tuning the ``sleep'' time $t_{\textrm{dead}}$ of the FPGA controlling the OS, in case of gated SPADs or dark-count-free detectors we can always have both detectors ready (i.e. out of their dead time) for each accepted heralding count, thus $N_{1}^{\mathrm{(Trig)}}=N_{2}^{\mathrm{(Trig)}}=N^{\mathrm{(Trig)}}$.
With free-running detectors, instead, $N^{\mathrm{(Trig)}}$ is just the number of heralding signals accepted by the FPGA (indicating the number of times the OS has been open), and a dead-time correction must be introduced (unless the dark counts are negligible).\\
Note that we have made the assumption that all events represented by $P$'s are mutually exclusive and independent of each other.
This is a reasonable assumption in the situation where these probabilities are $\ll1$, and the HBT SPADs settings are tuned in order to make afterpulses indistinguishable from dark counts.
To evaluate the photon count and dark count probabilities, we look at the temporal histograms of the SPADs detection events (as in Fig.\ref{peak-in}).\\
The following analysis allows us to use the time-tagged counts from individual detectors in combination with a time-correlated measurement from both to calculate Glauber's normalized second-order autocorrelation function:
\begin{equation}\label{g2tau}
  g^{(2)}(\tau)=\lim_{t\rightarrow\infty}\frac{\left\langle I(t)I(t+\tau) \right\rangle}{\left\langle I(t)\right\rangle \left\langle I(t+\tau) \right\rangle},
\end{equation}
with $I(t)$ being the photon emission intensity at time $t$.
For our purpose, the most interesting case appears at zero time difference ($\tau=0$), since an ideal SPS should present $g^{(2)}(0)=0$, meaning that the source is producing anti-bunched single photons.
Hence, the smaller the $g^{(2)}(0)$, the better the performances in terms of single-photon emission.
Actually, what we evaluate for our HSPS is the parameter $\alpha$ (approximating the second order correlation function $g^{(2)}(0)$ for $P_{i}^{\mathrm{(Tot)}} \ll 1$) \cite{grangier1}:
\begin{equation}\label{alfa}
\alpha=\frac{P_{12}^{\mathrm{(Ph;Ph)}}}{P_{1}^{\mathrm{(Ph)}} \cdot P_{2}^{\mathrm{(Ph)}}}\simeq g^{(2)}(0),
\end{equation}
where $P_{12}^{\mathrm{(Ph;Ph)}}$ is the probability of a coincidence photon count between the two HBT SPADs (dark counts subtracted).
Since $P_{12}^{\mathrm{(Her;Her)}}=0$ (since we have a probability below $10^{-9}$ of getting more than one heralded photon within a 1 ns time interval, which is a generous estimate of the detector jitter time), and $P_{i}^{\mathrm{(Bkg)}}$ is independent from $P_{i}^{\mathrm{(Dark)}}$, one has:
\begin{equation}\label{P12}
P_{12}^{\mathrm{(Ph;Ph)}}=P_{12}^{\mathrm{(Tot;Tot)}}-P_{1}^{\mathrm{(Tot)}}P_{2}^{\mathrm{(Dark)}}-
P_{1}^{\mathrm{(Dark)}}P_{2}^{\mathrm{(Tot)}}+P_{1}^{\mathrm{(Dark)}}P_{2}^{\mathrm{(Dark)}}.
\end{equation}
For each HBT SPAD, the quantity $N_{i}^{\mathrm{(Ph)}}$ is extracted as the difference between the counts of the photon peak region and a corresponding number of time bins in the dark count region, $N_{i}^{\mathrm{(Ph)}}=N_{i}^{\mathrm{(peak)}}-N_{i}^{\mathrm{(dark=peak)}}$.
From this, we can compute $N_{i}^{\mathrm{(Dark)}}=N_{i}^{\mathrm{(Tot)}}-N_{i}^{\mathrm{(Ph)}}$, and then we have all we need to evaluate the single-count probabilities in Eq. \ref{P12}.
The total coincidence probability $P_{12}^{\mathrm{(Tot;Tot)}}=\frac{N^{\mathrm{(Coinc)}}}{N^{\mathrm{(Trig)}}}$, being $N^\mathrm{(Coinc)}$ the number of two-photon counts events in the HBT, is independently evaluated by the coincidence electronics associated to it.\\
A further refinement for the $\alpha$ parameter evaluation arises when looking at the multichannel picture in Fig. \ref{peak-in}: it is in fact evident that the portion of dark counts on the left of the photon peak plays a different role with respect to the rest of them, since they occur before the arrival of the heralded photons.
Indeed, this portion of dark counts forbids the SPAD to detect the heralded photons.
This implies that such events (that we will indicate with $N_{i}^{\mathrm{(Null)}}$) should be removed from the valid trigger events as well as from the SPAD counts, and thus the true valid trigger counts, dark counts and total counts will be, respectively: ${N'}_{i}^{\mathrm{(Trig)}}=N^{\mathrm{(Trig)}}-N_{i}^{\mathrm{(Null)}}$, ${N'}_{i}^{\mathrm{(Dark)}}=N_{i}^{\mathrm{(Dark)}}-N_{i}^{\mathrm{(Null)}}$, ${N'}_{i}^{\mathrm{(Tot)}}=N_{i}^{\mathrm{(Tot)}}-N_{i}^{\mathrm{(Null)}}$.\\
This will affect all the different count probabilities, and thus the new formula for the $\alpha$ parameter will be:
\begin{equation}\label{alfa1}
\alpha=\frac{{P'}_{12}^{\mathrm{(Ph;Ph)}}}{{P'}_{1}^{\mathrm{(Ph)}} \cdot {P'}_{2}^{\mathrm{(Ph)}}}.
\end{equation}
%
By introducing the coefficient $q_i=N_{i}^{\mathrm{(Trig)}}/(N_{i}^{\mathrm{(Trig)}}-N_{i}^{\mathrm{(Null)}})$, one can write ${P'}_{i}^{\mathrm{(Ph)}}=q_i {P}_{1}^{\mathrm{(Ph)}}$. Analogously, one can derive for the coincidence count probability the relation ${P'}_{12}^{\mathrm{(Ph;Ph)}}=q_1q_2{P}_{12}^{\mathrm{(Ph;Ph)}}$.
This means that for the $\alpha$ parameter one has:
\begin{equation}\label{alfa2}
\alpha=\frac{{P'}_{12}^{\mathrm{(Ph;Ph)}}}{{P'}_{1}^{\mathrm{(Ph)}} \cdot {P'}_{2}^{\mathrm{(Ph)}}} = \frac{q_1q_2{P}_{12}^{\mathrm{(Ph;Ph)}}}{q_1{P}_{1}^{\mathrm{(Ph)}} \cdot q_2{P}_{2}^{\mathrm{(Ph)}}} = \frac{{P}_{12}^{\mathrm{(Ph;Ph)}}}{{P}_{1}^{\mathrm{(Ph)}} \cdot {P}_{2}^{\mathrm{(Ph)}}}.
\end{equation}
that shows how such correction, even though sensible for both single and coincidence count probabilities, does not affect $\alpha$.\\

\section{The HBT interferometers}

INRiM's HBT hosts a 50:50 telecom fiber beam splitter and two calibrated Micro Photon Devices Single Photon Counters based on InGaAs/InP SPADs, operating in gated mode with a $30$ ns detection window.
The two SPADs outputs, together with the FPGA gating signal triggering them, are addressed to a time-tagging coincidence electronics (PicoQuant HydraHarp400) with a 2.5 ps time-bin resolution as well as to a time-to-amplitude converter (TAC) module, in order to have a separate evaluation of the single counts and coincidence counts of the HBT SPADs.\\
Rearranging Eq. (\ref{alfa}) gives:
\begin{equation}\label{alfa3}
\alpha=\frac{P_{12}^{\mathrm{(Tot;Tot)}}-P_{1}^{\mathrm{(Tot)}}P_{2}^{\mathrm{(Tot)}} \left(Q_1^{\mathrm{(Dark)}}+Q_2^{\mathrm{(Dark)}}-Q_1^{\mathrm{(Dark)}}Q_2^{\mathrm{(Dark)}}\right)} {(P_{1}^{\mathrm{(Tot)}}(1-Q_{1}^{\mathrm{(Dark)}}))\cdot (P_{2}^{\mathrm{(Tot)}}(1-Q_{2}^{\mathrm{(Dark)}}))},
\end{equation}
being $P_{i}^{\mathrm{(Tot)}}=\frac{\left\langle\tilde{N}_{i}^{\mathrm{(Tot)}}\right\rangle}{\left\langle\tilde{N}^{\mathrm{(Trig)}}\right\rangle}$ and $Q_i^{\mathrm{(Dark)}}=\frac{N_i^{\mathrm{(Dark)}}}{N_i^{\mathrm{(Tot)}}}= \frac{N_i^{\mathrm{(Tot)}}-(N_i^{\mathrm{(peak)}}-N_i^{\mathrm{(dark=peak)}})}{N_i^{\mathrm{(Tot)}}}$, respectively, the total count probability and the fraction of dark counts registered by the $i$-th SPAD of the HBT.
The $Q_i^{\mathrm{(Dark)}}$ quantities are extracted by the MCA histograms of the two detectors forming the HBT, integrating all the counts registered by the time-tagging coincidence electronics within the whole acquisition time.
On the other hand, the $P_i^{\mathrm{(Tot)}}$ and $P_{12}^{\mathrm{(Tot;Tot)}}$ probabilities are obtained from the TAC counts, acquired in sets lasting 100 s each (the $\langle\tilde{N}\rangle$ quantities are the mean values of these repeated acquisitions).\\
NPL HBT hosts two different Id Quantique detectors, a gated SPAD (id210, with a 25 ns detection window) and a low-noise free-running SPAD (id230).
The SPADs outputs, together with the FPGA valid gate output (i.e. the heralding counts validated by the FPGA and used for the HBT SPADs gating), are routed to the HydraHarp400 time-tagger.
A first 1000 s acquisition is run feeding the HydraHarp400 with the HBT SPADs output going to the HydraHarp 400 channels 1 and 2, respectively, while the FPGA valid gate output is used as reference external clock.
Then, a second 2000 s run is instead performed with the id210 as the external clock and the id230 as channel 1 input.
Finally a third 1000 s acquisition is collected with the same settings as the first one.
This method gives the possibility of making a time-correlated evaluation of the HBT coincidence counts (during the 2000 s acquisition), whilst minimising the (eventual) temporal fluctuations and drifts of the photon source under test.
Since one of the HBT detectors is free-running, we need to set a sensible collection time window on the HydraHarp400; to be sure not to loose any significant events, we choose a 200 ns collection time window both for single- and two-photon events acquisitions.\\
Here, the total single-photon count probabilities will be $P_{i}^{\mathrm{(Tot)}}=\frac{N_i^{\mathrm{(Tot)}}}{N_S^{\mathrm{(Trig)}}}$, with $N_S^{\mathrm{(Trig)}}$ the number of FPGA valid gates belonging to the single-photon event acquisitions, while the two-photon count probability will be $P_{12}^{\mathrm{(Tot;Tot)}}=\frac{N^{\mathrm{(Coinc)}}}{N_C^{\mathrm{(Trig)}}}$, being $N_C^{\mathrm{(Trig)}}$ the FPGA valid gates related to the two-photon events acquisition.
The $Q_i^{\mathrm{(Dark)}}$ estimation remains instead unchanged.\\
PTB HBT hosts two Id Quantique gated SPADs (one id210 and one id201, with, respectively, 25 ns and 50 ns detection windows), both triggered by the same FPGA valid gate signal.
Since the time-tagging system chosen is a PicoQuant PicoHarp300, hosting only one input channel plus an external clock, the simultaneous evaluation of the single- and two-photon counts is not feasible.
For this reason, we use the same method as the second INRiM-NPL measurement session, with a first 1000 s acquisition in which the single counts of both detectors at once were evaluated (by feeding the joined output signals of the HBT SPADs to the PicoHarp300 input channel, adding a proper time delay between the two to allow discrimination, and using the FPGA valid gates as external clock), a second 2000 s acquisition with the id210 output as clock and the id201 output as signal to estimate the two-photon counts, and a third 1000 s acquisition with the same settings as the first one.\\
Concerning the data analysis, we adopt the same method as the INRIM-NPL session for the evaluation of $\alpha$ and the related uncertainty.

\section{Results}

Table \ref{tab1} hosts the results obtained by INRiM, PTB and NPL during the different measurement sessions, namely, the concurrent measurements of the $g^{(2)}(0)$ of the HSPS carried out firstly by INRiM and NPL and then by INRiM and PTB.\\
\begin{table}[htbp]
\begin{center}
\begin{tabular}{|c|c|c|c|}
  \hline
  Session & INRiM & NPL & PTB \\ \hline
  INRiM-NPL & $0.013\pm0.008$ & $0.02\pm0.02$ & - \\ \hline
  INRiM-PTB & $0.016\pm0.006$ & - & $0.04\pm0.05$ \\ \hline
\end{tabular}
\end{center}
\caption{Experimental results of the pilot comparison, with a coverage factor $k=1$.}
\label{tab1}
\end{table}
The uncertainty budgets related to the INRiM measurements are resumed in tables \ref{budget_inrim2} and \ref{budget_inrim3}, showing the single uncertainties, sensitivity coefficients and contributions to the total uncertainty on $\alpha$ for each measurement session.\\
\begin{table}[htbp]
\begin{center}
\begin{tabular}{|c|c|c|c|c|}
  \hline
   & Quantity & Unc. & Sens. Coeff. & Unc. Contr. \\
  \hline
  $Q_1^{\mathrm{(Dark)}}$ & $0.05604$ & $0.00008$ & $1.0461$ & $0.00010$ \\
  \hline
  $Q_2^{\mathrm{(Dark)}}$ & $0.05607$ & $0.00008$ & $1.0462$ & $0.00010$ \\
  \hline
  $\tilde{N}^{\mathrm{(Trig)}}$ & $6.0133\times10^6$ & $2.6\times10^3$ & $2.242\times10^{-8}$ & $0.00006$ \\
  \hline
  $\tilde{N}_1^{\mathrm{(Tot)}}$ & $18261$ & $31$ & $7.382\times10^{-6}$ & 0.0002 \\
  \hline
  $\tilde{N}_2^{\mathrm{(Tot)}}$ & $19396$ & $32$ & $6.950\times10^{-6}$ & $0.0002$ \\
  \hline
  $\tilde{N}^{\mathrm{(Coinc)}}$ & $7.1$ & $0.4$ & $0.01905$ & $0.008$ \\
  \hline\hline
  $\alpha$ & $0.013$ & 0.008\\
  \cline{1-3}
\end{tabular}
\end{center}
\caption{INRiM uncertainty budget related to the INRiM-NPL measurement session. Coverage factor $k=1$.}
\label{budget_inrim2}
\end{table}
\begin{table}[htbp]
\begin{center}
\begin{tabular}{|c|c|c|c|c|}
  \hline
   & Quantity & Unc. & Sens. Coeff. & Unc. Contr. \\
  \hline
  $Q_1^{\mathrm{(Dark)}}$ & $0.04525$ & $0.00008$ & $1.0302$ & $0.00008$ \\
  \hline
  $Q_2^{\mathrm{(Dark)}}$ & $0.04875$ & $0.00009$ & $1.0340$ & $0.00009$ \\
  \hline
  $\tilde{N}^{\mathrm{(Trig)}}$ & $6.1885\times10^6$ & $2.4\times10^3$ & $1.898\times10^{-8}$ & $0.00005$ \\
  \hline
  $\tilde{N}_1^{\mathrm{(Tot)}}$ & $22490$ & $41$ & $5.223\times10^{-6}$ & $0.0002$ \\
  \hline
  $\tilde{N}_2^{\mathrm{(Tot)}}$ & $23407$ & $43$ & $5.018\times10^{-6}$ & $0.0002$ \\
  \hline
  $\tilde{N}^{\mathrm{(Coinc)}}$ & $9.1$ & $0.5$ & $0.01294$ & $0.006$ \\
  \hline\hline
  $\alpha$ & $0.016$ & 0.006\\
  \cline{1-3}
\end{tabular}
\end{center}
\caption{INRiM uncertainty budget related to the INRiM-PTB measurement session. Coverage factor $k=1$.}
\label{budget_inrim3}
\end{table}
The uncertainty budget of the INRiM-NPL and INRiM-PTB measurement sessions are instead given in table \ref{budget_NPL2} and table \ref{budget_PTB1}, respectively.
Since here we do not have repeated measurements providing information on the photon counts fluctuations (in our regime, dark counts obey a Poisson distribution), we give an upper bound for them considering a superpoissonian behavior by putting $u(N)=\xi\sqrt{N}$, with $\xi>1$.\\
\begin{table}[htbp]
\begin{center}
\begin{tabular}{|c|c|c|c|c|}
  \hline
   & Quantity & Unc. & Sens. Coeff. & Unc. Contr. \\
  \hline
  $N_S^{\mathrm{(Trig)}}$ & $1.20348\times10^8$ & $\xi\cdot1.1\times10^4$ & $1.042\times10^{-9}$ & $\xi\cdot0.00002$ \\
  \hline
  $N_1^{\mathrm{(Tot)}}$ & $304900$ & $\xi\cdot600$ & $1.726\times10^{-7}$ & $\xi\cdot0.00010$ \\
  \hline
  $N_2^{\mathrm{(Tot)}}$ & $283300$ & $\xi\cdot600$ & $1.003\times10^{-7}$ & $\xi\cdot0.00005$ \\
  \hline
  $N_1^{\mathrm{(Dark)}}$ & $3100$ & $60$ & $3.256\times10^{-6}$ & $0.00018$ \\
  \hline
  $N_2^{\mathrm{(Dark)}}$ & $9600$ & $100$ & $3.590\times10^{-6}$ & $0.0004$ \\
  \hline
  $N_C^{\mathrm{(Trig)}}$ & $1.20184\times10^8$ & $\xi\cdot1.1\times10^4$ & $5.218\times10^{-10}$ & $\xi\cdot0.00001$ \\
  \hline
  $N^{\mathrm{(Coinc)}}$ & $43$ & $\xi\cdot7$ & $0.00146$ & $\xi\cdot0.010$ \\
  \hline\hline
  $\alpha$ & $0.02$ & 0.02\\
  \cline{1-3}\cline{1-3}
\end{tabular}
\end{center}
\caption{Uncertainty budget related to the NPL measurement of the INRiM-NPL measurement session. Coverage factor $k=1$. For the evaluation of the global uncertainty on $\alpha$, we set $\xi=2$.}
\label{budget_NPL2}
\end{table}
\begin{table}[htbp]
\begin{center}
\begin{tabular}{|c|c|c|c|c|}
  \hline
   & Quantity & Unc. & Sens. Coeff. & Unc. Contr. \\
  \hline
  $N_S^{\mathrm{(Trig)}}$ & $1.23807\times10^8$ & $\xi\cdot1.1\times10^4$ & $1.031\times10^{-8}$ & $\xi\cdot0.00011$ \\
  \hline
  $N_1^{\mathrm{(Tot)}}$ & $453500$ & $\xi\cdot700$ & $1.140\times10^{-6}$ & $\xi\cdot0.0008$ \\
  \hline
  $N_2^{\mathrm{(Tot)}}$ & $474200$ & $\xi\cdot700$ & $4.884\times10^{-7}$ & $\xi\cdot0.0004$ \\
  \hline
  $N_1^{\mathrm{(Dark)}}$ & $50700$ & $300$ & $2.390\times10^{-6}$ & $0.0006$ \\
  \hline
  $N_2^{\mathrm{(Dark)}}$ & $140800$ & $400$ & $2.888\times10^{-6}$ & $0.0011$ \\
  \hline
  $N_C^{\mathrm{(Trig)}}$ & $1.23733\times10^8$ & $\xi\cdot1.1\times10^4$ & $5.158\times10^{-10}$ & $\xi\cdot0.00006$ \\
  \hline
  $N^{\mathrm{(Coinc)}}$ & $690$ & $\xi\cdot30$ & $0.0009224$ & $\xi\cdot0.03$ \\
  \hline\hline
  $\alpha$ & $0.04$ & 0.05\\
  \cline{1-3}\cline{1-3}
\end{tabular}
\end{center}
\caption{Uncertainty budget related to the PTB measurement of the INRiM-PTB measurement session. Coverage factor $k=1$. For the evaluation of the global uncertainty on $\alpha$, we set $\xi=2$.}
\label{budget_PTB1}
\end{table}

\section{Conclusions}

We have illustrated a strategy to properly evaluate the multi-photon component of a CW light source, applied to a low-noise prototype of a fiber HSPS $@1550$ nm particularly suited for metrological and quantum-communication-related purposes and adaptable to a large variety of detectors and other devices.
The results of the whole measurement campaign, carried out with different measurement setups and data collection methodologies, are all in agreement within the experimental uncertainties, even with coverage factor $k=1$.\\
The proposed strategy may pave the way to a standardization of the characterisation of single-photon sources, a task of the utmost relevance for present and future metrology for quantum technologies.\\

\appendix

\section{Uncertainties evaluation}


\subsection{INRiM uncertainty budgets}

For what concerns the uncertainty budget related to INRiM measurements, we can identify three different contributions:
\begin{itemize}
  \item $u_{(P)}(\alpha)$: uncertainty derived from the double ($P_{12}^{\mathrm{(Tot;Tot)}}$) and single ($P_{i}^{\mathrm{(Tot)}}$) photon count probabilities, given by the TAC counts;
  \item $u_{(Q_i)}(\alpha)$, $i=1,2$: uncertainty derived from the $Q_i^{\mathrm{(Dark)}}$ fractions of each SPAD of the HBT, evaluated from the HydraHarp400 histograms.
\end{itemize}
The first uncertainty contribution, i.e. $u_{(P)}(\alpha)$, is evaluated in the following way:
\begin{equation}\label{uP}
u_{(P)}(\alpha)=\sqrt{\sum_{l,m=0}^3C_{l,m}\left(\frac{\partial\alpha} {\partial \tilde{N}_l}\right)\left(\frac{\partial\alpha} {\partial \tilde{N}_m}\right) u(\tilde{N}_l)u(\tilde{N}_m)}
\;\;\;\;\;
C_{l,m}=\frac{\langle \tilde{N}_l \tilde{N}_m \rangle - \langle \tilde{N}_l \rangle\langle \tilde{N}_m \rangle}{u(\tilde{N}_l)u(\tilde{N}_m)}
\end{equation}
being $\tilde{N}_0=\tilde{N}^{\mathrm{(Trig)}}$, $\tilde{N}_i=\tilde{N}_i^{\mathrm{(Tot)}}\;(i=1,2)$, $\tilde{N}_3=\tilde{N}^{\mathrm{(Coinc)}}$ and $u(\tilde{N}_l)$ the standard deviation on the average $\langle\tilde{N}_l\rangle$.
Table \ref{budget_inrim5} hosts the correlation coefficients $C_{i,j}$ ($i,j=0,...,3$) among the $\tilde{N}_i$'s for all of the INRiM measurement sessions.\\
\begin{table}[htbp]
\begin{center}
\begin{tabular}{|c|c|c|}
  \hline
   & INRiM-NPL & INRiM-PTB \\
  \hline
  $C_{0,1}$ & $0.326$ & $0.857$ \\
  \hline
  $C_{0,2}$ & $0.445$ & $0.891$ \\
  \hline
  $C_{0,3}$ & $0.269$ & $0.133$ \\
  \hline
  $C_{1,2}$ & $0.650$ & $0.824$ \\
  \hline
  $C_{1,3}$ & $0.00261$ & $-0.000155$ \\
  \hline
  $C_{2,3}$ & $-0.000909$ & $0.00161$ \\
  \hline
\end{tabular}
\end{center}
\caption{Correlation coefficients related to the INRiM measurements in both sessions.}
\label{budget_inrim5}
\end{table}
Concerning the uncertainties contributions $u_{(Q_i)}(\alpha)$, we have to consider that extracting them just from the HydraHarp400 histograms would lead to an overestimation of the total uncertainty $u(\alpha)$; on the contrary, knowing that $Q_i^{\mathrm{(Dark)}}=\frac{N_i^{\mathrm{(Dark)}}}{N_i^{\mathrm{(Tot)}}}=1-\frac{\tilde{N}_i^{\mathrm{(Ph)}}}{\tilde{N}_i^{\mathrm{(Tot)}}}$, we can write:
\begin{equation*}
  u^2{(Q_i^{\mathrm{(Dark)}})}= \left(\frac{1}{\tilde{N}_i^{\mathrm{(Tot)}}}\right)^2 u^2(\tilde{N}_i^{\mathrm{(Ph)}}) + \left(\frac{\tilde{N}_i^{\mathrm{(Ph)}}}{(\tilde{N}_i^{\mathrm{(Tot)}})^2}\right)^2 u^2(\tilde{N}_i^{\mathrm{(Tot)}}) -
\end{equation*}
\begin{equation}\label{uQi1}
   2\left(\frac{1}{\tilde{N}_i^{\mathrm{(Tot)}}}\right)\left(\frac{\tilde{N}_i^{\mathrm{(Ph)}}}{(\tilde{N}_i^{\mathrm{(Tot)}})^2}\right) \left( \langle \tilde{N}_i^{\mathrm{(Tot)}}\tilde{N}_i^{\mathrm{(Ph)}}\rangle - \langle \tilde{N}_i^{\mathrm{(Tot)}}\rangle\langle \tilde{N}_i^{\mathrm{(Ph)}}\rangle \right).
\end{equation}
Since $\tilde{N}_i^{\mathrm{(Tot)}}=\tilde{N}_i^{\mathrm{(Ph)}}+\tilde{N}_i^{\mathrm{(Dark)}}$, with some algebraic passages one has:
\begin{equation}\label{uQi2a}
\left( \langle \tilde{N}_i^{\mathrm{(Tot)}}\tilde{N}_i^{\mathrm{(Ph)}}\rangle - \langle\tilde{N}_i^{\mathrm{(Tot)}}\rangle\langle\tilde{N}_i^{\mathrm{(Ph)}}\rangle \right)= u^2(\tilde{N}_i^{\mathrm{(Ph)}}) \leq u^2(\tilde{N}_i^{\mathrm{(Tot)}}),
\end{equation}
and hence:
\begin{equation}\label{uQi2}
u{(Q_i^{\mathrm{(Dark)}})} \leq \frac{Q_i^{\mathrm{(Dark)}}}{\tilde{N}_i^{\mathrm{(Tot)}}}u(\tilde{N}_i^{\mathrm{(Tot)}}),
\end{equation}
giving a sensible upper bound on the uncertainty contributions $u_{(Q_i)}(\alpha)=\left|\frac{\partial\alpha}{\partial Q_i^{\mathrm{(Dark)}}}\right|u{(Q_i^{\mathrm{(Dark)}})}$.
Finally, as no correlation exists among these three contributions, the total uncertainty on $\alpha$ is:
\begin{equation}\label{u_tot}
u(\alpha)=\sqrt{u^2_{(Q_1)}(\alpha)+u^2_{(Q_2)}(\alpha)+u^2_{(P)}(\alpha)}
\end{equation}

\subsection{NPL and PTB uncertainty budgets}

Concerning NPL and PTB measurements, for evaluating the uncertainty $u(\alpha)$ we use:
\begin{equation}\label{uNPL}
u(\alpha)=\sqrt{\sum_{l=0}^6\left(\frac{\partial\alpha} {\partial N_l}\right)^2 u^2(N_l)},
\end{equation}
with $N_0=N_S^{\mathrm{(Trig)}}$, $N_i=N_i^{\mathrm{(Tot)}}$ and $N_{2+i}=N_i^{\mathrm{(Dark)}}$ ($i=1,2$), $N_5=N_C^{\mathrm{(Trig)}}$ and, finally, $N_6=N^{\mathrm{(Coinc)}}$.\\
Since here we don't have repeated measurements, we have to provide a sensible upper bound to the measurement uncertainties by statistical considerations on the physics of the source and measurement device.
The dark counts of the HBT SPADs are known to follow a Poisson distribution.
The statistics of a multi-mode PDC, like the one exploited in our HSPS, is almost indistinguishable from a Poissonian distribution.
For these reasons, we can assume the fluctuations of the $N_l$ quantities in (\ref{uNPL}) to be Poissonian as well, and choose as related uncertainties $u(N_l)=\xi\sqrt{N_l}$, being $\xi$ a coefficient to be tuned taking into account the (eventual) superpoissonian behavior of the system.
In our case, having observed some unexpected fluctuations in the CW pump power, to be conservative we choose $\xi=2$. \\

\end{document}